\documentclass{ws-ijmpb}

\usepackage{psfig}
\usepackage{amstext}

\def\draftnote{}

\begin{document}

  \markboth{M.~Kollar}{Construction of a Dispersion Relation}

  \title{Construction of a dispersion relation from an
    arbitrary density of states}

  \author{\footnotesize MARCUS KOLLAR}
  
  \address{In\-sti\-tut f\"{u}r Theore\-ti\-sche Phy\-sik,\\
    Jo\-hann~Wolf\-gang~Goethe-Uni\-ver\-si\-t\"{a}t
    Frank\-furt,\\
    Ro\-bert-Mayer-Stra\ss{}e~8-10, D-60054~Frankfurt am~Main,
    Germany\\
  }

  \maketitle
  
  \pub{~}{~}

  \begin{abstract}
    The dispersion relations of energy bands in solids are
    characterized by their density of states, but a given density of
    states may originate from various band structures. We show how a
    spherically symmetric dispersion can be constructed for any
    one-band density of states.  This method is applied to one-, two-
    and three-dimensional systems.  It also serves to establish that
    any one-band spectrum with finite bandwidth can be obtained from a
    properly scaled dispersion relation in the limit of infinite
    dimensions.
  \end{abstract}

  \section{Introduction}

  The band structure of electrons describes their
  quan\-tum-mechan\-i\-cal motion in the periodic potential of a
  crystal lattice in the absence of interactions.  This picture is
  modified by the Cou\-lomb interaction between them, which must be
  taken into account accurately to understand the phenomena observed
  in strongly correlated systems such as transition-metal oxides,
  ferromagnetic metals, or the cuprate superconductors.  Nonetheless
  the properties of the non-interacting lattice system remain an
  important input quantity.
  
  The electronic spectrum of a solid is characterized by the density
  of states,
  \begin{eqnarray}
    N(\omega)
    =
    \frac{1}{V_{\text{B}}}
    \int\!\textrm{d}k_1
    \!\cdot\!\cdot\!
    \int\!\textrm{d}k_D
    \;
    \delta\!\left(\omega-\epsilon_{\vec{k}}\right)
    \!,\label{eq:N-of-omega}
  \end{eqnarray}
  where the integral is over the first Brillouin zone, which has
  volume $V_{\text{B}}$; here and in the following we consider only a
  single separated band $\epsilon_{\vec{k}}$.  The mapping from
  dispersion relation to density of states is not invertible, i.e.,
  two different dispersions may lead to the same function $N(\omega)$.
  Nevertheless in certain situations it is desirable to construct a
  dispersion that corresponds to a particular $N(\omega)$.  In this
  paper we present a method that performs this inversion in any
  dimension $D$, with additional assumptions on the functional form of
  $\epsilon_{\vec{k}}$.
  
  Construction of a dispersion with given spectral properties is
  especially of interest within the framework of dynamical mean-field
  theory (DMFT) for Hubbard-type models, which becomes exact in the
  limit of infinite dimensions, $D$ $\to$
  $\infty$.\cite{Metzner89a,Georges96a} In this approach typical
  tight-binding dispersions, arising from hopping of electrons between
  orbitals on neighboring lattice sites, lead to a density of states
  which extends to infinite values of $\omega$. For example, on the
  hypercubic lattice $N(\omega)$ becomes a Gaussian bell
  curve\cite{Metzner89a}
  \begin{eqnarray}
    N(\omega)
    =
    \frac{1}{\sqrt{2\pi}}
    \,
    \exp\!\left(
      -\frac{\omega^2}{2}
    \right)
    \!.\label{eq:N-gaussian}    
  \end{eqnarray}
  Exponential tails are also obtained for the generalized
  face-centered cubic lattice\cite{MuellerHartmann91a} and the
  generalized diamond lattice.\cite{Santoro93a}
  
  The DMFT is particularly well suited to describe the Mott-Hubbard
  metal-insulator transition.\cite{Gebhard97a} In order to work with a
  finite bandwidth, which better models finite-dimensional systems,
  DMFT studies of this transition (for recent work see
  Ref.~\refcite{Bulla01a} and references therein) have concentrated on
  next-neigh\-bor hopping on the Bethe lattice in the limit of
  infinite connectivity.  It leads to a semi-circular density of
  states,\cite{Georges96a}
  \begin{eqnarray}
    N(\omega)
    =
    \frac{2}{\pi}
    \,
    \sqrt{1-\omega^2}
    \,,\;\;\;
    |\omega|\leq1
    \,,\label{eq:N-bethe}
  \end{eqnarray}
  which in this case is defined via the one-particle Green's function,
  since there is no underlying periodic crystal lattice and hence no
  dispersion relation.  The Bethe lattice has hitherto been the only
  known infinite-dimensional system with a finite bandwidth. Below we
  will show how a dispersion can be constructed on a $D$-dimensional
  crystal lattice that in the limit $D$ $\to$ $\infty$ yields any
  given density of states with finite bandwidth, providing an answer
  to one of the open conceptual questions about this limit.
  
  In DMFT all one-particle quantities in the homogeneous phase involve
  the dispersion only implicitly through $N(\omega)$. Therefore
  another standard approach has been to use a function $N(\omega)$
  that derives from a dispersion on a finite-dimensional lattice or is
  tuned by hand.  For instance, Wahle {\it et al.}\cite{Wahle98a}
  introduced the model density of states,
  \begin{eqnarray}
    N(\omega)
    =
    \frac{1+\sqrt{1-a^2}}{\pi}
    \,
    \frac{\sqrt{1-\omega^2}}{1+a\omega}
    \,,\;\;\;
    |\omega|\leq1
    \,,\label{eq:N-tunable}
  \end{eqnarray}
  to study the effect of asymmetry in $N(\omega)$ on the stability of
  ferromagnetism in the Hubbard model.  Here $a\in{[-1;1]}$ is a
  tunable parameter; (\ref{eq:N-tunable}) reduces to the symmetric
  function (\ref{eq:N-bethe}) for $a$ $=$ $0$. For $a>0$ spectral
  weight is shifted towards the lower band edge, while for $a$ $=$ $1$
  $N(\omega)$ has a square-root singularity at $\omega$ $=$ $-1$
  similar to the tight-binding density of states for the fcc lattice.
  Below we will derive dispersions in $D$ $=$ $1,2,3,\infty$ that
  represent this density of states.

  \section{Isotropic Fermi surfaces on Bravais lattices}

  We construct a dispersion with a given density of states as follows.
  In terms of the latter the band filling $n(\mu)$ for Fermi energy
  $\mu$ is given by
  \begin{eqnarray}
    n(\mu)
    =
    \int\limits_{\omega_{\text{min}}}^{\mu}\!\textrm{d}\omega
    \;
    N(\omega)
    \,,\label{eq:n-of-mu}
  \end{eqnarray}
  which lies between 0 and 1. Suppose now that the dispersion relation
  is spherically symmetric, $\epsilon_{\vec{k}}$ $=$ $\epsilon(|\vec{k}|)$,
  with $|\vec{k}|^2$ $=$ $k_1+\cdots+k_D^2$.  Furthermore we assume that
  $\epsilon(k)$ increases with $k$; then the Fermi sea is simply
  connected and includes the point $\vec{k}$ $=$ $0$, and there is a
  one-to-one relation between Fermi energy $\mu$ and Fermi vector $k$.
  The band filling can thus be obtained in another, geometric way as
  the normalized volume of the Fermi sea, $v_D(k)$, defined by
  \begin{eqnarray}
    v_D(k)
    &=&
    \int\limits_0^k\!\textrm{d}k'
    \;
    s_D(k')
    \,,\label{eq:v-def}
  \end{eqnarray}
  where $s_D(k)$ is its normalized surface,
  \begin{eqnarray}
    s_D(k)
    &=&
    \frac{1}{V_{\text{B}}}
    \int\!\textrm{d}k_1
    \!\cdot\!\cdot\!
    \int\!\textrm{d}k_D
    \;
    \delta\!\left(k-\sqrt{k_1^2+\cdots+k_D^2}\right)
    \!.\label{eq:s-def}
  \end{eqnarray}
  The dispersion $\epsilon(k)$ is now defined by setting $v_D(k)$
  equal to the band filling calculated from the density of states,
  i.e.,
  \begin{eqnarray}
    v_D(k)
    &=&
    n(\epsilon(k))
    \,.\label{eq:connection}
  \end{eqnarray}
  As is appropriate for a one-band density of
  states we assume $N(\omega)>0$ for
  $\omega_{\text{min}}<\omega<\omega_{\text{max}}$, so that $n(\mu$)
  is invertible on this interval.  In terms of its inverse function
  $\mu(n)$ we can write
  \begin{eqnarray}
    \epsilon(k)
    =
    \mu(v_D(k))
    \,.\label{eq:epsilon-calculated}
  \end{eqnarray}
  Note that the functional form of $\mu(n)$ depends only on the given
  density of states $N(\omega)$, whereas that of $v_D(k)$ depends only
  on the crystal lattice. Once the latter has been calculated for a
  lattice of interest, $\epsilon(k)$ is immediately available for any
  $N(\omega)$.
  
  Similarly, the density of states can be at once calculated from a
  given spherically symmetric dispersion $\epsilon(k)$, provided
  $\epsilon'(k)>0$. In terms of $k(\epsilon)$, the inverse function of
  $\epsilon(k)$, we have $n(\epsilon)$ $=$ $v_D(k(\epsilon))$ from
  (\ref{eq:epsilon-calculated}), and hence
  \begin{eqnarray}
    N(\epsilon)
    =
    n'(\epsilon)
    =
    s_D(k(\epsilon))\,k'(\epsilon)
    \,.\label{eq:N-calculated}
  \end{eqnarray}
  To apply the method in either direction it remains to calculate the
  functions $s_D(k)$ and $v_D(k)$.

  \section{Dimensions $D$ $=$ $1$, $2$, $3$}

  We consider now the special case of a hypercubic lattice in $D$
  dimensions, treating explicitly the cases $D$ $=$ $1,2,3$, as well
  as the limit $D$ $\to$ $\infty$. For convenience we measure $k$ in
  units of $2\pi/a$, where $a$ is the lattice spacing, so that the
  first Brillouin zone is located in the region $|k_i|<{\frac{1}{2}}$,
  and the maximum value for $|\vec{k}|$ is $\sqrt{D}/2$.  For
  $k\leq{\frac{1}{2}}$ the integrations in (\ref{eq:s-def}) can be
  performed in spherical coordinates, since the vector $\vec{k}$ always
  remains within the first Brillouin zone. In this case we have
  $s_D(k)$ $=$ $S_D\,k^{D-1}$ and $v_D(k)$ $=$ $V_D\,k^{D}$, where
  $S_D$ $=$ $2\pi^{D/2}/\Gamma(D/2)$ and $V_D$ $=$
  $\pi^{D/2}/\Gamma(1+D/2)$ are the surface and volume of the
  $D$-dimensional unit sphere, respectively.  This case applies for
  all $k$ in $D$ $=$ $1$, so that
  \begin{eqnarray}
    s_1(k)
    =
    2
    \,,
    \;\;
    v_1(k)
    =
    2k
    \,,
    \;\;
    \text{~for~}
    k\in{[0;{\textstyle\frac{1}{2}}]}
    \,,
  \end{eqnarray}
  representing the contribution of the two Fermi points. The spherical
  symmetry reduces to inversion symmetry in this simple case, which
  was discussed already in Ref.~\refcite{Wahle98a}.
  
\begin{figure}[t]
    \centerline{\psfig{file=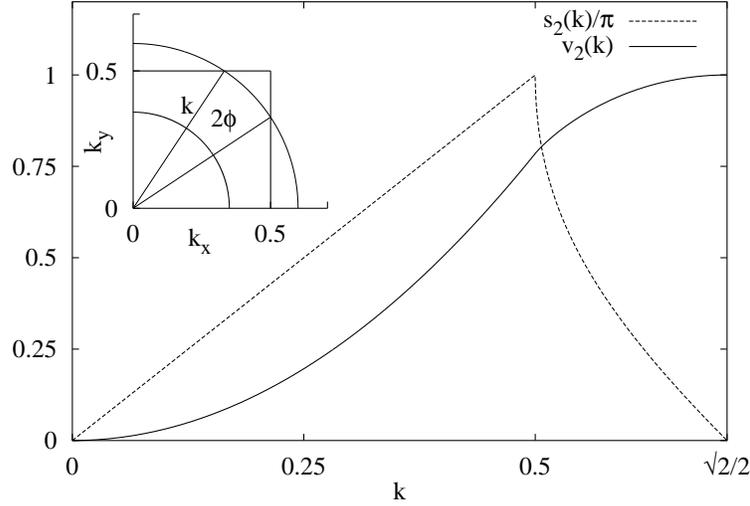,width=0.8\hsize}}
    \caption{Surface $s_2(k)$ and volume $v_2(k)$ of a circular Fermi
      sea with Fermi vector $k$ on the square lattice.  Inset: If the
      Fermi surface has radius $k\leq{\frac{1}{2}}$ it lies completely
      within the first Brillouin zone.  If $k>{\frac{1}{2}}$ the
      inside part contributes $2\phi k$ in each quadrant to $s_2(k)$,
      where $\cos\phi$ $=$ $1/(2k)$, leading to
      (\ref{eq:s2}).\label{fig:func2}}
  \end{figure}

  For the square lattice in $D$ $=$ $2$ the values $0\leq
  k\leq\sqrt{2}/2$ are allowed.  However if $k>{\frac{1}{2}}$ only
  that part of a circle with radius $k$ contributes which lies inside
  the square Brillouin zone, as shown in the inset of
  Figure~\ref{fig:func2}.  We obtain
  \begin{eqnarray}
    s_2(k)
    =
    \left\{
      \begin{array}{ll}
        2\pi k
        &
        \text{~if~}
        k\in{[0;\frac{1}{2}]}
        \\
        2\pi k-8k\arccos\frac{1}{2k}
        &
        \text{~if~}
        k\in{[\frac{1}{2};\frac{\sqrt{2}}{2}]}
      \end{array}
    \right.
    \!,\label{eq:s2}
  \end{eqnarray}
  and by integrating with respect to $k$,
  \begin{eqnarray}
    v_2(k)
    =
    \left\{
      \begin{array}{ll}
        \pi k^2
        &
        \text{~if~}
        k\in{[0;\frac{1}{2}]}
        \\
        \pi k^2
        +
        \sqrt{4k^2-1}
        -
        4k^2\arccos\frac{1}{2k}
        &
        \text{~if~}
        k\in{[\frac{1}{2};\frac{\sqrt{2}}{2}]}
      \end{array}
    \right.
    \!.\label{eq:v2}
  \end{eqnarray}
  These functions are shown in Figure~\ref{fig:func2}.
  
\begin{figure}[t]
    \centerline{\psfig{file=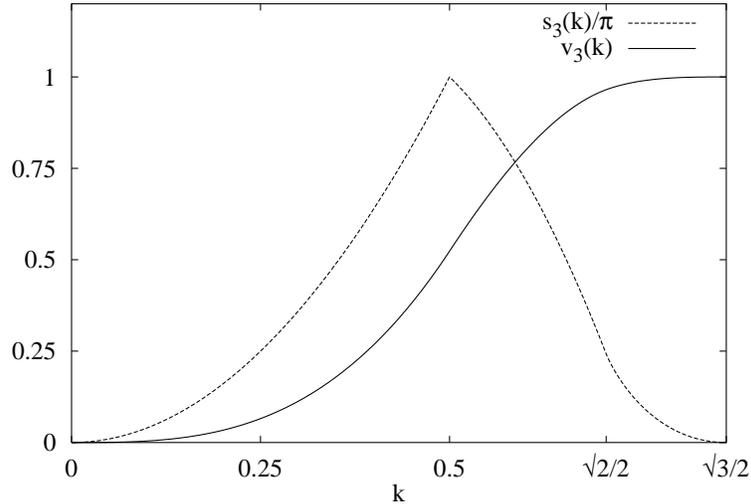,width=0.8\hsize}}
    \caption{Surface $s_3(k)$ and volume $v_3(k)$ of a spherical Fermi
      sea with Fermi vector $k$ on the simple cubic
      lattice.\label{fig:func3}}
  \end{figure}

  In $D$ $=$ $3$ the Brillouin zone of the simple cubic lattice allows
  the values $0\leq k\leq\sqrt{3}/2$, and more cases must be
  distinguished in the calculation of $s_3(k)$.  We omit geometric
  considerations and instead employ the recursion formula for $s_D(k)$
  on the hypercubic lattice,
  \begin{eqnarray}
    s_{D+1}(k)
    &=&
    \int\limits_{
      \text{max}(0,k^2-\frac{1}{4})^{\frac{1}{2}}
      }^{\text{min}(k,\frac{\sqrt{D}}{2})}
    \!\!\!\!
    \textrm{d}q
    \;\;
    \frac{2k\,s_D(q)}{\sqrt{k^2-q^2}}
    \,,
  \end{eqnarray}
  which can be obtained in a straightforward fashion from
  (\ref{eq:s-def}). Applying it to the above expression for $s_2(k)$ we
  obtain, after a lengthy calculation,
  \begin{eqnarray}
    s_3(k)
    =
    \left\{
      \begin{array}{ll}
        4\pi k^2
        &
        \text{~if~}
        k\in{[0;\frac{1}{2}]}
        \\
        6\pi k-8\pi k^2
        &
        \text{~if~}
        k\in{[\frac{1}{2};\frac{\sqrt{2}}{2}]}
        \\
        6\pi k-8\pi k^2+48k^2\arctan{\sqrt{1-\frac{1}{2k^2}}}
        \\ \;\;
        \,-\,
        24k\arctan{\sqrt{4k^2-2}}
        &
        \text{~if~}
        k\in{[\frac{\sqrt{2}}{2};\frac{\sqrt{3}}{2}]}
      \end{array}
    \right.
    \!,\label{eq:s3}
  \end{eqnarray}
  and one more integration yields
  \begin{eqnarray}
    v_3(k)
    =
    \left\{
      \begin{array}{ll}
        \frac{4}{3}\pi k^3
        &
        \text{~if~}
        k\in{[0;\frac{1}{2}]}
        \\
        3\pi k^2-\frac{8}{3}\pi k^3-\frac{\pi}{4}
        &
        \text{~if~}
        k\in{[\frac{1}{2};\frac{\sqrt{2}}{2}]}
        \\
        3\pi k^2-\frac{8}{3}\pi k^3-\frac{\pi}{4}
        +16k^3\arctan{\sqrt{1-\frac{1}{2k^2}}}
        \\ \;\;
        \,+\,
        \sqrt{4k^2-2}
        -
        (12k^2-1)\arctan{\sqrt{4k^2-2}}
        &
        \text{~if~}
        k\in{[\frac{\sqrt{2}}{2};\frac{\sqrt{3}}{2}]}
      \end{array}
    \right.
    \!.\label{eq:v3}
  \end{eqnarray}
  These functions are displayed in Figure~\ref{fig:func3}.

\begin{figure}[t]
    \centerline{\psfig{file=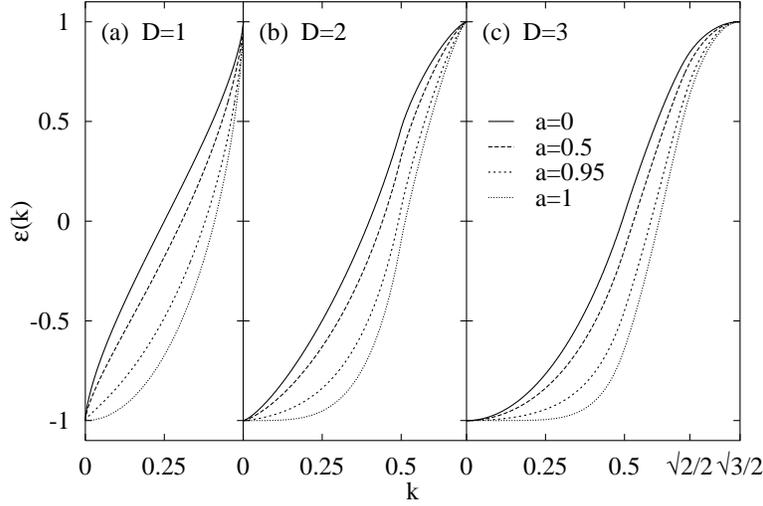,width=0.8\hsize}}
    \caption{Dispersions in $D$ $=$ $1,2,3$ leading to the
      model density of states (\ref{eq:N-tunable}) for several
      values of the asymmetry parameter $a$.\label{fig:disp}}
  \end{figure}

  As a simple example, let us consider the rectangular (constant)
  density of states
  \begin{eqnarray}
    N(\omega)
    =
    1
    \,,\;\;\;
    0\leq\omega\leq1
    \,,\label{eq:N-rectangular} 
  \end{eqnarray}
  where $\mu(n)$ $=$ $n$, and hence $\epsilon(k)$ $=$ $v_D(k)$.  Thus
  the plot of $v_D(k)$ in Figures~\ref{fig:func2} and \ref{fig:func3}
  also represents the shape of the dispersion that will yield the
  density of states (\ref{eq:N-rectangular}).  Next we apply the
  method to the model density of states (\ref{eq:N-tunable}). The band
  filling is now obtained as
  \begin{eqnarray}    
    n(\mu)
    &=&
    1
    -
    \frac{1+\sqrt{1-a^2}}{\pi a^2}
    \Bigg(\arccos\mu-a\sqrt{1-\mu^2}
    \nonumber\\&&~~~~
    \,-\,
    2\sqrt{1-a^2}
    \arctan\!\Bigg[\frac{(1-a)(1-\mu)}{(1+a)(1+\mu)}\Bigg]^{\frac{1}{2}}
    \;\Bigg)
    \,,\label{eq:tunable-filling}
  \end{eqnarray}
  which for the Bethe lattice ($a=0$) reduces to $n(\mu)$ $=$
  ${\frac{1}{2}}+{[\mu\sqrt{1-\mu^2}+\arcsin\mu]}/\pi$.
  The resulting dispersion relations are shown in
  Figure~\ref{fig:disp}.  Their curvature depends strongly on the
  asymmetry parameter $a$, and also on $D$.
  
  Another quantity of interest is the hopping amplitude between two
  lattice sites connected by a lattice vector $\vec{R}$ $=$
  $(R_1,\ldots R_D)$ with integer components. It is defined as (note
  the conventional overall minus sign)
  \begin{eqnarray}
    t(\vec{R})
    =-
    \int\limits_{-\frac{1}{2}}^{\frac{1}{2}}\!\textrm{d}k_1
    \!\cdot\!\cdot\!
    \int\limits_{-\frac{1}{2}}^{\frac{1}{2}}\!\textrm{d}k_D
    \;
    \epsilon({\textstyle\sqrt{k_1^2+\cdots+k_D^2}})
    \;
    \textrm{e}^{2\pi \textrm{i}\vec{k}\cdot\vec{R}}
    \,.\label{eq:t-def}
  \end{eqnarray}
  We calculate the hopping amplitude (\ref{eq:t-def}) for the
  one-dimensional dispersion (Figure~\ref{fig:disp}a).  For small $R$
  we find that the amplitudes for hopping between nearest and between
  next-nearest neighbors, $t(1)$ and $t(2)$, are of opposite sign, and
  $|t(2)/t(1)|$ is appreciable for $a$ close to $1$; for example
  $t(1)$ $=$ $0.337$, $t(2)$ $=$ $-0.080$, for $a$ $=$ $0.95$.  For
  large $|R|$ it can be shown\cite{vanDongen} that
  $t(R)\propto(-1)^R\,|R|^{-5/3}$. While the stabilization of the
  ferromagnetic phase was attributed to large spectral weight below
  the Fermi energy in Ref.~\refcite{Wahle98a}, the present calculation
  shows that from a complementary point of view (\ref{eq:N-tunable})
  can be regarded approximately as a $t$-$t'$ model, which has been
  shown to be favorable for ferromagnetism by other
  methods.\cite{Daul98a}

  \section{Limit of large dimensions}
  
  Finally we turn to the limit $D$ $\to$ $\infty$. To obtain a
  factorized expression for $s_D(k)$ for the hypercubic lattice we
  transform the argument of the delta function and use the
  representation $\delta(x)$ $=$
  $\int_{-\infty}^{\infty}\!\textrm{d}t\,
  \textrm{e}^{\textrm{i}tx}/(2\pi)$,
  \begin{eqnarray}
    s_D(k)
    &=&
    2k
    \int\limits_{-\frac{1}{2}}^{\frac{1}{2}}\!\textrm{d}k_1
    \!\cdot\!\cdot\!
    \int\limits_{-\frac{1}{2}}^{\frac{1}{2}}\!\textrm{d}k_D
    \;
    \delta\!\left(k_1^2+\cdots+k_D^2-k^2\right)
    \!,\label{eq:s-step1}
    \\
    &=&
    \frac{k}{\pi}
    \int\limits_{-\infty}^{\infty}\!\textrm{d}t
    \;
    \textrm{e}^{-\textrm{i}tk^2}
    \;
    g_0(t)^D
    \,,\label{eq:s-step2}
  \end{eqnarray}
  where we defined $g_0(t)$ $=$
  $\int_{-1/2}^{1/2}\textrm{d}q\,\textrm{e}^{\textrm{i}tq^2}$.  The
  maximum value for $k$ is $\sqrt{D}/2$; it is useful to introduce
  scaled quantities, $\kappa$ $=$ $2k/\sqrt{D}$ and
  $\bar{s}_D(\kappa)$ $=$ $s_D(\kappa\sqrt{D}/2)\sqrt{D}/2$,
  $\bar{v}_D(\kappa)$ $=$ $v_D(\kappa\sqrt{D}/2)$, with the property
  $\int_0^1\!\textrm{d}\kappa\,\bar{s}_D(\kappa)$ $=$ $\bar{v}_D(1)$
  $=$ $1$.  We change variables to $s$ $=$ $tD/4$,
  \begin{eqnarray}
    \bar{s}_D(\kappa)
    =
    \frac{\kappa}{\pi}
    \int\limits_{-\infty}^{\infty}\!\textrm{d}s
    \;
    \textrm{e}^{-\textrm{i}s\kappa^2}
    \;
    g_0({\textstyle\frac{4s}{D}})^D
    \,.\label{eq:sbar-D}
  \end{eqnarray}
  This equation is valid for all $D$.  From the expansion $g_0(t)$ $=$
  $1+\textrm{i}t/12-t^2/160+\textrm{O}(t^3)$ we obtain for large $D$
  \begin{eqnarray}
    g_0({\textstyle\frac{4s}{D}})^D
    =
    \exp\!\left(
      \frac{\textrm{i}s}{3}-\frac{s^2}{10D}
    \right)
    (1+\textrm{O}({\textstyle\frac{1}{D^2}}))
    \,,\label{eq:g0-infty}
  \end{eqnarray}
  which, when substituted into (\ref{eq:sbar-D}), leads to a sharply
  peaked Gaussian for large $D$,
  \begin{eqnarray}
    \bar{s}_D(\kappa)
    =
    \frac{2\kappa}{\sqrt{2\pi}\,\sigma}
    \exp\!\left(
      -\frac{(\kappa^2-\frac{1}{3})^2}{2\sigma^2}
    \right)
    (1+\textrm{O}({\textstyle\frac{1}{D^2}}))
    \,,\label{eq:sbar-infty}
  \end{eqnarray}
  with the abbreviation $\sigma$ $=$ $1/\sqrt{5D}$. We find that for
  large $D$ the function $\bar{s}_D(\kappa)$ eventually approaches a
  delta function, $\bar{s}_D(\kappa)\sim\delta(\kappa-1/\sqrt{3})$ and
  $s_D(k)\sim\delta(k-\sqrt{D/12})$. This surprisingly simple result
  means that in this limit the Fermi sea is essentially located in a
  narrow hyperspherical shell (cut off at the Brillouin zone
  boundaries) at roughly half the maximum value of $k$. Interestingly,
  the Fermi surface has the same average radius as for the
  tight-binding dispersion with nearest-neighbor
  hopping.\cite{MuellerHartmann89b} The Fermi surface in the latter
  case is distributed with a Gaussian in $\kappa-1/\sqrt{3}$ instead
  of $\kappa^2-1/3$; these distributions become equivalent for $D$
  $\to$ $\infty$.
  
\begin{figure}[t]
    \centerline{\psfig{file=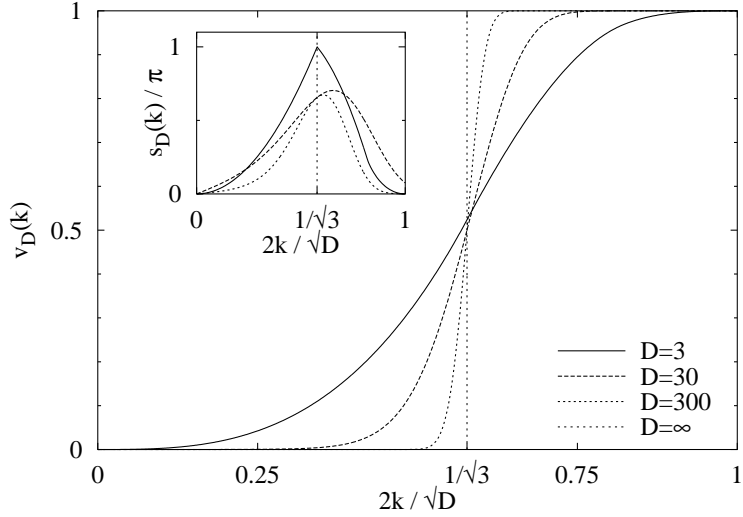,width=0.8\hsize}}
    \caption{Weight functions $s_D(k)$ and $v_D(k)$,
      approaching (\ref{eq:sbar-infty}) and (\ref{eq:vbar-infty}) for
      large dimensions $D$. For $D$ $=$ $3$ the closed forms
      (\ref{eq:s3}) and (\ref{eq:v3}) are shown.  For example, in
      dimension $D$ the dispersion $\epsilon(k)=v_D(k)$ will yield the
      rectangular density of states
      (\ref{eq:N-rectangular}).\label{fig:infty}}
  \end{figure}

  Next we obtain the asymptotic expression for $\bar{v}_D(\kappa)$
  from (\ref{eq:sbar-infty}) to the same order,
  \begin{eqnarray}
    \bar{v}_D(\kappa)
    =
    \left\{
      \begin{array}{ll}
        \frac{1}{2}
        \left[
          \text{erfc}(\frac{1-3\kappa^2}{3\sqrt{2}\sigma})
          -
          \text{erfc}(\frac{1     }{3\sqrt{2}\sigma})
        \right]\!
        &
        \text{~if~}
        \kappa\in{[0;\frac{1}{\sqrt{3}}]}
        \\[2mm]
        1-
        \frac{1}{2}
        \left[
          \text{erfc}(\frac{3\kappa^2-1}{3\sqrt{2}\sigma})
          +
          \text{erfc}(\frac{1          }{3\sqrt{2}\sigma})
        \right]\!
        &
        \text{~if~}
        \kappa\in{[\frac{1}{\sqrt{3}};1]}
      \end{array}
    \right.
    \!.\label{eq:vbar-infty}
  \end{eqnarray}
  This function ultimately approaches a step function,
  $\bar{v}_D(\kappa)\sim\Theta(\kappa-1/\sqrt{3})$ and
  $v_D(k)\sim\Theta(k-\sqrt{D/12})$; however further calculations must
  be performed with the asymptotic expressions (\ref{eq:sbar-infty})
  and (\ref{eq:vbar-infty}) instead of the delta and step functions.
  In particular (\ref{eq:epsilon-calculated}) and
  (\ref{eq:vbar-infty}) prescribe how to construct a dispersion that
  in the limit of infinite dimensions will yield any desired
  $N(\omega)$ with finite bandwidth, for example (\ref{eq:N-bethe}) or
  (\ref{eq:N-tunable}).  The functions $\bar{v}_D(\kappa)$ and
  $\bar{s}_D(\kappa)$ are shown in Figure~\ref{fig:infty}.  As $D$
  approaches infinity, the dispersion thus has the form of a
  smoothed-out step function, changed in shape by the application of
  $\mu(n)$.
  
  Let us consider some examples.  As noted above the plot of $v_D(k)$
  in Figure~\ref{fig:infty} already shows the dispersion that yields
  the rectangular density of states (\ref{eq:N-rectangular}).  On the
  other hand, for the special quadratic dispersion
  \begin{eqnarray}
    \epsilon(k)
    =
    \sqrt{5D}
    \left(
      \frac{4k^2}{D}-\frac{1}{3}
    \right)
    \!,\label{eq:epsilon-quadratic} 
  \end{eqnarray}
  we use (\ref{eq:sbar-infty}) and (\ref{eq:N-calculated}) to obtain
  the corresponding density of states
  \begin{eqnarray}
    N(\omega)
    =
    \frac{1}{\sqrt{2\pi}}
    \,
    \exp\!\left(
      -\frac{\omega^2}{2}
    \right)
    ,\;\;\;
    -\sqrt{5D}/3\leq\omega\leq2\sqrt{5D}/3
    \,,\label{eq:N-quadratic}
  \end{eqnarray}
  to leading order in $1/D$. Hence in the limit $D$ $\to$ $\infty$ the
  dispersion (\ref{eq:epsilon-quadratic}) yields the Gaussian
  (\ref{eq:N-gaussian}). This shows that the tight-binding spectrum
  for next-neighbor hopping on the hypercubic lattice may
  alternatively be represented by a quadratic dispersion in the limit
  $D$ $\to$ $\infty$.
  
  It is instructive to calculate the hopping amplitude
  (\ref{eq:t-def}) corresponding to the dispersion
  (\ref{eq:epsilon-calculated}) in the limit of large $D$. We change
  variables to $|\vec{k}|$, introduce the scaled dispersion
  $\bar{\epsilon}(\kappa)$ $=$ $\epsilon(\kappa\sqrt{D}/2)$, and
  perform a similar factorization as before, with the result (valid
  for all $D$)
  \begin{eqnarray}
    t(\vec{R})
    =-
    \int\limits_{0}^{1}\!\textrm{d}\kappa
    \,
    \frac{\kappa\,\bar{\epsilon}(\kappa)}{\pi}
    \int\limits_{-\infty}^{\infty}\!\textrm{d}s
    \,
    \textrm{e}^{-\textrm{i}s\kappa^2}
    \,
    \prod_{j=1}^D
    g_{R_j}({\textstyle\frac{4s}{D}})^D
    \,,\label{eq:t-D}
  \end{eqnarray}
  where we introduced $g_R(t)$ $=$
  $\int_{-1/2}^{1/2}dq\,\textrm{e}^{\textrm{i}tq^2+2\pi
    \textrm{i}qR}$.  Note that by setting $\vec{R}$ $=$ $0$ in
  (\ref{eq:t-D}) the center of mass of the band, $t(0)$ $=$
  $\int_{\omega_{\text{min}}}^{\omega_{\text{max}}}\!\textrm{d}\omega
  \,\omega\,N(\omega)$, is correctly recovered.  Proceeding to the
  limit $D$ $\to$ $\infty$ we first observe that $g_R(t)$ $=$
  $-\textrm{i}t(-1)^R/(2\pi^2R^2)+\textrm{O}(t^2)$ for integer
  $R\neq0$. The hopping amplitude along a coordinate axis is then
  calculated to leading order in $1/D$ as
  \begin{eqnarray}
    t_1(R_1)
    &\equiv&
    t((R_1,0,\ldots,0))
    \nonumber\\
    &=&
    \frac{(-1)^{R_1}}{D\pi^2R_1^2}
    \int\limits_{0}^{1}\!\textrm{d}\kappa
    \,
    \frac{\kappa\,\bar{\epsilon}(\kappa)}{\pi}
    \int\limits_{-\infty}^{\infty}\!\textrm{d}s
    \,
    2\textrm{i}s
    \,
    \exp\!\left[
      \textrm{i}s
      \!\left(
        \frac{1}{3}-\kappa^2
      \right)\!
      -\frac{s^2}{10D}
      \right]\!
    \nonumber\\
    &=&
    \frac{(-1)^{R_1}}{D\pi^2R_1^2}
    \int\limits_{0}^{1}\!\textrm{d}\kappa
    \;
    \bar{\epsilon}(\kappa)
    \,
    \frac{d}{d\kappa}
    \left(
      -\frac{\bar{s}_D(\kappa)}{\kappa}
    \right)
    \!.\label{eq:t1}
  \end{eqnarray}
  Using (\ref{eq:sbar-infty}) the remaining integral is calculated by
  partial integration and yields
  $\sqrt{3}\,\bar{\epsilon}\,'(1/\sqrt{3})$, which is evaluated via
  (\ref{eq:epsilon-calculated}),
  \begin{eqnarray}
    \bar{\epsilon}\,'(\kappa)
    =
    \mu'(\bar{v}_D(\kappa))\,\bar{s}_D(\kappa)
    =
    \frac{\bar{s}_D(\kappa)}{N(\mu(\bar{v}_D(\kappa)))}
    \,.\label{eq:epsilon-prime}
  \end{eqnarray}
  To leading order in $1/D$ we have $\bar{s}_D(1/\sqrt{3})$ $=$
  $\sqrt{10D/3\pi}$ and $\bar{v}_D(1/\sqrt{3})$ $=$ ${\frac{1}{2}}$,
  whence
  \begin{eqnarray}
    t_1(R_1)
    =
    \frac{t_1^{\ast}}{\sqrt{D}}
    \;
    \frac{(-1)^{R_1}}{R_1^2}
    \,,\;\;\;
    t_1^{\ast}
    =
    \frac{\sqrt{10}}{\pi^{\frac{5}{2}}N(\mu(\frac{1}{2}))}
    \,.\label{eq:t1-result}
  \end{eqnarray}
  The numerical factor $t_1^{\ast}$ contains the value of the density
  of states at the Fermi energy for a half-filled band,
  $\mu({\frac{1}{2}})$.  For fixed $R_1$ this type of hopping takes
  place from each site to $2D$ other lattice sites, and the hopping
  amplitude is scaled proportional to the inverse square root of this
  coordination number, which is also the required scaling for a
  tight-binding band with hopping to nearest
  neighbors.\cite{Metzner89a} An important difference is that in the
  present case the hopping is long ranged and slowly decaying. It is
  also apparent that $t(\vec{R})$ has higher symmetry than the crystal
  lattice, which is a consequence of the spherical symmetry of
  $\epsilon_{\vec{k}}$.  This spherical symmetry is also responsible
  for the large amplitude for long-range hopping, as any type of
  finite-range hopping would lead to a dispersion relation with only a
  discrete symmetry.
  
  In similar fashion we calculate the hopping amplitude between two
  sites con\-nected by a lattice vector with two nonzero entries,
  $t_2(R_1,R_2)\equiv t((R_1,R_2,0,$ $\ldots,0))$,
  \begin{eqnarray}
    t_2(R_1,R_2)
    =
    \frac{t_2^{\ast}}{D}
    \;
    \frac{(-1)^{R_1+R_2}}{R_1^2R_2^2}
    \,,\;\;\;
    t_2^{\ast}
    =
    \frac{10N'(\mu(\frac{1}{2}))}{\pi^5N(\mu(\frac{1}{2}))^3}
    \,.\label{eq:t2-result}
  \end{eqnarray}
  The value of $t_2^{\ast}$ now also involves the first derivative of
  the density of states; for example, $t_2^{\ast}=-0.080$ and
  $t_1^{\ast}=0.839$ for (\ref{eq:N-tunable}) with $a=0.95$, but
  $t_2^{\ast}=0$ for (\ref{eq:N-bethe}).  The number of sites
  connected by hopping of type (\ref{eq:t2-result}) is proportional to
  $D^2$, the square root of which again appears in the scaling factor
  as expected. We conclude that although each hopping amplitude is
  scaled with dimension in the same way as in the tight-binding case,
  the present long-range hopping amplitudes combine to yield an energy
  band with a finite width, in contrast to the infinite tails in the
  tight-binding spectrum.
  
  Finally, note that the nesting symmetry, $\epsilon_{\vec{k}}$ $=$
  $-\epsilon_{\vec{k}+\vec{Q}}$, which results if hopping takes place
  only between different sublattices of a bipartite lattice, is absent
  for a spherically symmetric dispersion.  Here $\vec{k}+\vec{Q}$ is
  taken modulo the first Brillouin zone and the antiferromagnetic wave
  vector is $\vec{Q}$ $=$
  $(\frac{1}{2},\frac{1}{2},\ldots,\frac{1}{2})$ in our notation. For
  example, the spherically symmetric representation
  (\ref{eq:epsilon-quadratic}) of the Gaussian density of states
  yields
  \begin{eqnarray}
    \epsilon_{\vec{k}+\vec{Q}}
    =
    \sqrt{5D}
    \left(
      \frac{4k^2}{D}+\frac{2}{3}-\frac{4}{D}\sum_{i=1}^D|k_i|
    \right)
    \!\label{eq:quadratic-nesting} 
  \end{eqnarray}
  for $|k_i|\leq\frac{1}{2}$.  Hence $\epsilon_{\vec{k}}$ $\neq$
  $-\epsilon_{\vec{k}+\vec{Q}}$ for most of the Brillouin zone, in
  particular on or near the Fermi surface, where
  $|k_i|=\frac{1}{2\sqrt{3}}$ $+$ $O(D^{-1/2})$.  That the nesting
  condition is not fulfilled can also be seen from the fact that sites
  on the same sublattice are connected by the long-range hopping
  amplitude (\ref{eq:t1-result}).  In general, this frustration is
  expected to suppress antiferromagnetism in the Hubbard model, which
  is of interest in particular for the study of the Mott-Hubbard
  metal-insulator transition in the paramagnetic phase.  It seems that
  spherically symmetric dispersion relations offer a conceptually and
  calculationally simpler way to remove the antiferromagnetic phase
  than, e.g., ``fully frustrated'' hopping on the Bethe
  lattice.\cite{Georges96a}

  \section{Conclusion and outlook}
  
  In conclusion we presented a method that, for any crystal lattice,
  maps a one-band density of states to a dispersion
  $\epsilon_{\vec{k}}$ that only depends on $|\vec{k}|$.  This
  procedure unphysically enlarges the usual invariance of the
  dispersion under the point group of the crystal to a continuous
  spherical symmetry.  As a consequence, the nesting symmetry typical
  for nearest-neighbor hopping on bipartite lattices is absent.
  Furthermore, the dispersion may have kinks if the specified density
  of states has van Hove singularities that are inappropriate for the
  dimension $D$ of the representing lattice.  On the other hand, the
  spherical symmetry may allow for analytic calculations of DMFT
  one-particle quantities in the homogeneous phase that depend on the
  dispersion only through the density of states, such as the Green
  function of the Hubbard model in the vicinity of the Mott-Hubbard
  metal-insulator transition.
  
  \section*{Acknowledgements}
  
  The author would like to thank S.~Sachdev and D.~Vollhardt for
  valuable discussions. During a stay at Yale Unversity this research
  was supported by US NSF Grant DMR 0098226.

\end{document}